# Strain Control of Oxygen Vacancies in Epitaxial Strontium Cobaltite Films

*Jonathan R. Petrie[1], Chandrima Mitra[1], Hyoungjeen Jeen[1], Woo Seok Choi[1], Tricia L. Meyer[1], Fernando A. Reboredo[1], John W. Freeland[2], Gyula Eres[1], and Ho Nyung Lee[1*]*

[1]Materials Science and Technology Division, Oak Ridge National Laboratory, Oak Ridge, TN, 37831, USA.

[2]Advanced Photon Source, Argonne National Laboratory, Argonne, IL, 60439, USA.

E-mail: hnlee@ornl.gov

**The ability to manipulate oxygen anion defects rather than metal cations in complex oxides is facilitating new functionalities critical for emerging energy and device technologies. However, the difficulty in activating oxygen at reduced temperatures hinders the deliberate control of an important defect, oxygen vacancies. Here, strontium cobaltite ($SrCoO_x$) is used to demonstrate that epitaxial strain is a powerful tool for manipulating the oxygen vacancy concentration even under highly oxidizing environments and at annealing temperatures as low as 300 °C. By applying a small biaxial tensile strain (2%), the oxygen activation energy barrier decreases by ~30%, resulting in a tunable oxygen deficient steady-state under conditions that would normally fully oxidize unstrained cobaltite. These strain-induced changes in oxygen stoichiometry drive the cobaltite from a ferromagnetic metal towards an antiferromagnetic insulator. The ability to decouple the oxygen vacancy concentration from its typical dependence on the operational environment is useful for effectively designing oxides materials with a specific oxygen stoichiometry.**

# 1. Introduction

Originally seen as undesirable and detrimental to the performance of functional transition metal oxide (TMOs) materials, the formation of oxygen vacancies is becoming increasingly important due to the realization that these same defects can lead to new functional phenomena.[1-3] For instance, incremental changes in oxygen vacancies can leverage large shifts in magnetic, electronic, and catalytic properties in TMOs without introducing possible impurities and segregation associated with heterovalent cation doping.[4-7] Moreover, the functional manipulation of oxygen vacancies is critical for several key information, energy, and environmental technologies, including high $T_c$ superconductors, colossal magnetoresistive materials, oxygen membranes, energy storage, memristors, and other electrochemical devices.[8-12] Traditionally, the concentration and ordering of these vacancies have been dictated by the exposure of oxides to a reducing environment, e.g. oxygen partial pressure and temperature. However, this passive reliance on the environment severely limits the functionality of oxygen vacancies in electrochemical devices, such as batteries and solid oxide fuel cells (SOFCs) that operate at reduced temperatures (< 600 °C) far from the optimal conditions for vacancy activation.[13, 14] To tailor the content of oxygen vacancies to the increasing number of novel applications that utilize these functional defects, a new method of controlling the oxygen defect concentration is required.

Epitaxial strain in thin film heterostructures is known to critically affect a multitude of physical properties, such as magnetic ordering, electron mobility, ferroelectricity, and superconductivity.[15-19] Recent computational studies have investigated the effects of epitaxial strain on oxygen vacancies in conventional perovskites. While they predict that the activation energies of these vacancies can be tuned by tenths of an eV under modest strains of a few percent, the initial energies of >2 eV render these changes relatively insignificant at reduced temperatures.[20-22] Therefore, to use strain as a new parameter for tuning the oxygen vacancy concentration, oxides with lower activation energies are in high demand.

In this regard, strontium cobaltite, $SrCoO_x$ (SCO), has sparked interest due to the discovery of a low-temperature topotactic transition between the brownmillerite phase $SrCoO_{2.5}$, denoted as BM-SCO, and the perovskite phase $SrCoO_{3-\delta}$, denoted as P-SCO, where $0 \leq \delta \leq 0.25$.[23-25] Due to both the easy intercalation of $O^{2-}$ within BM-SCO offered by ordered vacancy channels (OVCs) and the metastability of $Co^{4+}$ in P-SCO, the cobaltite has exceptionally low oxygen activation energies (< 1 eV), amplifying the energetic effects of strain.[26] In addition, deviations in oxygen content from the near-stoichiometric P-SCO result in significant property changes from a ferromagnetic metal to an antiferromagnetic insulator.[23] Furthermore, since the fast, reversible redox reactions offered by these materials promote the $Co^{3+}/Co^{4+}$ redox couple, epitaxial SCO oxygen sponges have revealed enhanced catalytic activities towards CO oxidation at ~300 °C, making these films attractive for sensors and energy-related devices, such as SOFCs[23, 27] The combination of unique properties and such low energetic thresholds for oxygen control made SCO films an ideal platform for systematically studying strain-induced oxygen non-stoichiometry in these oxides and its ability to influence physical properties.

In this paper, we report that modest amounts of epitaxial strain, specifically tensile strain, can significantly reduce the oxygen activation energy, which effectively leads to a tunable oxygen deficiency even in a highly oxidizing annealing environment. By approaching such an oxygen deficiency from either a topotactically transformed BM-SCO or fully-oxidized P-SCO film, the non-stoichiometry is shown as the equilibrium state. This variation in oxygen stoichiometry under conditions that would normally allow only stoichiometric perovskites results in tailored and systematic shifts in magnetic and electrical transport properties, demonstrating the control of functionality afforded by strain engineering in perovskite-based oxides.

## 2. Results and Discussion

### 2.1. Structural Analysis of Annealed BM-SCO and P-SCO Films

To explore the effects of oxidation on either BM-SCO or fully oxidized, nearly stoichiometric perovskite directly grown in ozone (P-SCO$_{ozone}$)[23], films were epitaxially deposited on lattice-mismatched substrates using pulsed laser epitaxy (PLE), as detailed in the Experimental section. All films had uniform thicknesses of 15 nm to ensure no strain relaxation on various perovskite substrates. The substrates included (001) (LaAlO$_3$)$_{0.3}$-(SrAl$_{0.5}$Ta$_{0.5}$O$_3$)$_{0.7}$ (LSAT), (001) SrTiO$_3$ (STO), (110) DyScO$_3$ (DSO), (110) GdScO$_3$ (GSO), and (001) KTaO$_3$ (KTO). Their pseudo-cubic parameters range from $a_{sub}$ = 3.868 to 3.989 Å (see **Figure 1**). While BM-SCO is orthorhombic ($a_o$ = 5.574, $b_o$ = 5.447, $c_o$ = 15.745 Å), it will be represented throughout this paper as pseudo-tetragonal ($a_t$ = 3.905, $c_t/4$ = 3.936 Å).[24] The nearly stoichiometric P-SCO$_{ozone}$, on the other hand, is cubic with $a_c$ = 3.829 Å, leading to substrate-induced lattice mismatches from 1.0 to 4.2%, as shown in Figure 1. Both BM-SCO and P-SCO$_{ozone}$ films were subsequently annealed *in-situ* at 300 ºC in 500 Torr of O$_2$ for 5 minutes to explore the effects of strain in a highly oxidizing atmosphere known to convert unstrained SCO to the nearly stoichiometric perovskite.[23, 25] As was reported elsewhere,[20,22] such annealing time and environment is known to fully oxidize samples. In either case, oxygen in BM-SCO or P-SCO$_{ozone}$ is respectively intercalated or deintercalated through the OVCs to approach a thermodynamically stable P-SCO state, where stoichiometry under even high oxygen partial pressure is dependent on strain (see Figure 1). XRD reciprocal space mapping confirmed that brownmillerite and perovskite films used here were coherently strained (see **Figure S1** and **Figure S2**).

Figure 1a shows an example of our observation that strain can modulate oxygen stoichiometry in epitaxial P-SCO films that have undergone topotactic oxidation. The XRD $\theta$-$2\theta$ scans around the P-SCO 002 peak for annealed BM-SCO on various substrates are given. The $\theta$-$2\theta$ scans for annealed P-SCO$_{ozone}$ are virtually identical. Each perovskite peak is clearly defined with Kiessig fringes that verify the superior film quality. However, upon careful inspection of the out-of-plane lattice constant, the monotonic shift in the out-of-plane lattice parameter with tensile strain cannot be simply understood through a Poisson-type contraction due to substrate-induced tensile strain. Therefore, we further considered the possibility of lattice expansion due to increased vacancy formation as the film deviates from the stoichiometric $\delta = 0$ P-SCO phase in SrCoO$_{3-\delta}$.[22, 28] We compared the unit cell volume of these annealed films to that of the as-grown P-SCO$_{ozone}$ films (see **Figure S3** for $\theta$-$2\theta$ scans), which have essentially stoichiometric concentrations of oxygen.[23, 29, 30] As seen in Fig. 1b, the unit cell volume of the P-SCO$_{ozone}$ films increased monotonically with strain for all but the $\varepsilon = 4.2\%$ film, which may be due to a slight change in oxygen stoichiometry.[31] This increase was readily fit to a line ascribing all linear expansion to a Poisson ratio of $v \sim 0.26$, which is a common value associated with cobaltites.[32] Similarly, the BM-SCO had a $v \sim 0.30$, as seen in **Figure S4**. However, the topotactically oxidized P-SCO films exhibited a larger lattice volume than that of as-grown P-SCO$_{ozone}$. This deviation becomes more pronounced when the tensile strain increases. As summarized in Fig. 1b, this increased unit cell volume is observed from P-SCO samples prepared by either annealing BM-SCO or P-SCO$_{ozone}$, indicating a steady-state vacancy concentration attained through, respectively, either the intercalation or deintercalation of oxygen. As the increase in the oxygen vacancies often results in lattice expansion for perovskite-typed complex oxides, we primarily attribute this deviation to the greater oxygen deficiency in the films with tensile strain;[28] otherwise, an unrealistic $v = 0.17$ would be required from fully oxygenated films to fit the experimental data.[33]

**2.2. Oxygen Stoichiometry Changes probed by X-ray Absorption Spectroscopy**

To confirm that the oxygen stoichiometry in the film varies as the tensile strain is increased, we investigated the topotactically oxidized P-SCO films by x-ray absorption spectroscopy (XAS) using both the O *K*- and Co *L*-edges. As seen in Fig. 2a, there are two peaks of note in the pre-peak region of the O *K*-edge. These peaks are labeled A and B and are linked to Co 3*d* – O 2*p* hybridization from hole states associated, respectively, with either fully oxidized or partially oxidized coordination.[34] While the intensity of Peak B increases under tensile strain with oxygen loss, peak A substantially diminishes as less intercalated oxygen translates into an oxygen deficient state. Peak A also shifts to higher photon energies with oxygen loss as a result of negative charge-transfer.[35] The shift of 0.2 eV as strain increases to 4.2% is consistent with the transition from $SrCoO_{2.9}$ to $SrCoO_{2.75}$ determined in previous studies for bulk P-SCO.[36]

An investigation of the Co-*L* edge in Fig. 2b also indicates a changing valence state with increasing amounts of oxygen vacancies at higher tensile strains. The shift in intensity of the Co-$L_{2,3}$ peaks towards lower energies confirms that there is indeed a decrease in the average transition metal valency from $Co^{4+}$ with increasing strain. The chemical shift in the Co $L_2$ edge between the $\varepsilon = 1.0$ and 4.2% films is -0.4 eV. A related -1 eV shift in the Co $L_2$ edge can be seen in the $(La_{1-x}Sr_x)CoO_3$ system when one electron is transferred away from Co as the Sr concentration varies from $x = 1$ to $x = 0$.[37] Using this shift, we can estimate a transfer here of up to 0.4 e$^-$, which is compatible with $\Delta\delta \leq 0.20$ as the equilibrium state transitions from $SrCoO_{2.9}$ to SrCoO$_{2.75}$ with tensile strain. Quantification of the oxygen non-stoichiometry is given in **Figure 3**. The estimates of oxygen deficiency were obtained from comparing structural data (unit cell volume)[29, 30] and spectroscopic peak shifts (O-*K* and Co-*L* edges)[34, 37] to previous studies.

**2.2. Evolution of Magnetic and Electrical Transport Properties with Strain**

Varying the oxygen vacancy content through strain has powerful effects on the magnetic and electronic properties of the cobaltite. It is already theoretically predicted that strain can diminish the ferromagnetic and metallic states of stoichiometric P-SCO due to reduction in the orbital overlap.[38, 39] However, as seen in the magnetic and electrical transport properties in **Figure 4**, increasing oxygen non-stoichiometry complements strain in driving the transition from a ferromagnetic metal to an antiferromagnetic insulator. To decouple the effects of pure strain from the effects of strain-induced changes in the oxygen stoichiometry, magnetization and transport properties are shown for both the as-grown, fully-oxidized P-SCO$_{ozone}$ as well as the non-stoichiometric annealed P-SCO. Figure 4a and 4b show the respective magnetic moments of the P-SCO$_{ozone}$ and annealed P-SCO films as a function of magnetic field at 10 K. Due to the large magnetic moments of the paramagnetic DSO and GSO substrates, magnetization data of thin films on those substrates are not shown. As expected from theoretical studies, the saturation magnetization $M_s$ of the fully-oxidized P-SCO$_{ozone}$ slightly decreases after $\varepsilon = 2\%$ due to lessening of the orbital overlap.[38, 39] However, the annealed P-SCO shows a much more dramatic decrease in $M_s$, declining from 2.3 $\mu_B$/Co for the $\varepsilon = 1\%$ case to 0.1 $\mu_B$/Co for $\varepsilon =$ 4.2%. The $M_s$ as a function of temperature in Figure 4c and 4d also shows the weakening of the ferromagnetic properties from a Curie temperature, $T_c$, above 230 K for all strained P-SCO$_{ozone}$ to a range from ~220 to 160 K on annealed SCO, decreasing with strain. Furthermore, according to a recent bulk study [40] $T_c$ ~ 160 K, 220 K, and 280 K respectively indicated off-stoichiometric

P-SCO phases $SrCoO_{2.75}$, $SrCoO_{2.90}$, and $SrCoO_3$. Therefore, the reduction of $T_c$ in our annealed P-SCO thin films samples suggests a change from $δ \sim 0.1$ for $ε = 1\%$ to $δ \sim 0.25$ for $ε = 4.2\%$.

Moreover, electrical *dc* transport measurements in Figure 4e and 4f show a correlation with the changing oxygen stoichiometry due to strain. As seen from the electrical transport data in Figure 4e, P-SCO$_{ozone}$ films with strains of at most $ε = 2.0\%$ are metallic and indicate increasing insulating behavior with tensile strain due to increased localization of carriers.[38, 39] When comparing these films to the annealed P-SCO in Figure 4e, the annealed P-SCO films reveal slightly increased resistivities due to strain-induced changes in oxygen stoichiometry. In fact, the only substrate on which the annealed P-SCO film is metallic is LSAT, which induces a strain of only ~ 1% in the film. Since it has been shown that the double exchange mechanism thought responsible for the metallic behavior of P-SCO is disrupted when $δ \geq 0.1$, we can place an upper bound on $δ$ at this level when 1% tensile strain is supplied to the film.[36, 41] At mismatches that induce even greater tensile strains, the transition to an increasingly insulating nature implies that $δ > 0.1$ and that $δ$ is increasing with such strain. Consistent with the XAS and magnetic data, none of the films (even with strain) exhibit behaviors that imply $δ > 0.25$ or the BM-SCO phase; in comparison to the resistivity of a previously-deposited 14 nm BM-SCO film on STO, all annealed P-SCO films have values orders of magnitude lower up to room temperature. In addition, the derived activation energy from the transport measurements for conductive species increases to only 87 meV for P-SCO under 4.2% strain compared to 240 meV for the BM-SCO film over a similar temperature range. This again suggests that while the films are increasingly oxygen deficient, no tensile strain was sufficient to topotactically transform the P-SCO film back to BM-SCO under these conditions.

This electronic trend has also been observed in our prior study;[24] however, the origin for the less conducting behavior under tensile strain has not been completely understood. Indeed, while the systematic trend is obvious, the more insulating nature from the P-SCO films compared to P-SCO$_{ozone}$ indicates that their carrier transport is strongly influenced by the change in strain-induced carrier concentration owing to the loss of oxygen. However, an optical spectroscopy study (see **Figure S5**) revealed that even the highly insulating P-SCO film on STO maintains similar spectral features as the metallic P-SCO on LSAT, which has clear Drude features.[25] Thus, we conclude that the P-SCO thin films are at the verge of the percolation limit with a possible coexistence of metallic (P-SCO) and insulating (BM-SCO) phases. A similar trend was reported in manganite bulk and thin film samples, which also showed a discrepancy between optical and dc transport data[42].

## 2.3. Strain Control of Oxygen Activation Energy and Formation Enthalpy

To understand the experimental observation of the strong coupling between strain and oxygen stoichiometry, we performed first-principles Density Functional Theory (DFT) calculations.[26, 43-46] We specifically computed two quantities as a function of strain (see Supplementary Information for details). The first is the formation enthalpy, $H_i$, to intercalate an oxygen atom into one of the vacancy sites, which is the point of lowest energy in the system. The other is the activation energy barrier, $E_a$, which is the energy required for an oxygen atom to diffuse from one vacancy site to another along OVCs in the open network structure[47]. As seen in **Figure 5a**, in diffusing from one vacancy site to the other along the [010] direction in the channel, $O^{2-}$ crosses an intermediate saddle point, $H_{saddle}$, which is the point of highest energy in the system.[13] This results in $E_a$ being composed of the difference between $H_{saddle}$ and $H_i$. While $H_{saddle}$ only slightly lowers with tensile strain, $H_i$ significantly increases with tensile strain due to changes in the stabilizing effects of hybridization between Co 3$d$ and O 2$p$, which are dependent on the Co-O bond length[20].

As shown in Figure 5b, the net effect of such tensile strain reduces $E_a$, whereas it is raised by compressive strain. It is worth stressing that by applying only a 2% tensile strain, one can reduce the activation energy barrier by ~30%. Such declines immediately suggest benefits towards ionic conduction for SOFCs and oxygen sensor applications[2-4]. Concurrent with the strain-induced changes in $E_a$, as shown in Fig. 3b, $H_i$ rises with tensile strain and falls with compressive strain, respectively suggesting either greater or smaller thermodynamic oxygen instability. Both values indicate that the application of tensile strain significantly facilitates oxygen vacancy generation, whereas compressive strain prevents the system from losing oxygen, due to shifts in the enthalpy of the vacancy sites. By controlling the energetics of the vacancy sites through strain, shifts due to environmental influences can be minimized and oxygen vacancy dependent properties, such as conductivity and magnetism, adjusted in conditions that would fully oxidize the perovskite towards a simple ferromagnetic metal.

## 3. Conclusion

In summary, this work shows that by growing epitaxially-strained SrCoO$_{3-\delta}$ thin films we can tailor oxygen non-stoichiometry crucial for functionality simply by applying tensile strain to lower the equilibrium oxygen concentration. We attribute the phenomenon of easier oxygen loss in tensile-strained films to reduced oxygen activation energy from weakened Co-O bonding. Remarkably, this approach is capable of controlling oxygen vacancies even in highly oxidizing environments, producing large changes in magnetic and electronic properties suitable for sensors and devices designed to work in oxidizing conditions and reduced temperatures. With the discovery that tensile strain dictates the oxygen stoichiometry by controlling the activation energies in strontium cobaltites, deterministic control over physical and electrochemical properties can be envisioned. Thus, the discovery that tensile strain enhances the creation of oxygen defects opens up a new route for designing novel functional oxides using strain as the key tuning parameter.

## 4. Experimental Section

*Thin Film Synthesis*: Epitaxial films of BM-SCO and P-SCO were grown 15-nm thick on different substrates through pulsed laser epitaxy (PLE). Similar to previous work,[23-25] the BM-SCO growth temperature, oxygen partial pressure, laser fluence, and repetition rate were fixed at 750 °C, 100 mTorr, 1.5 J/cm$^2$, and 5 Hz, respectively. For annealing *in-situ*, the as-deposited BM-SCO films were cooled under 100 mTorr of $O_2$ to 300 °C before introducing 500 Torr $O_2$ into the chamber to topotactically oxidize the films for 5 minutes. P-SCO$_{ozone}$ films were grown under the same conditions with the exception of the partial pressure, which was 200 mTorr of a mix of $O_2+O_3$(5%). All films were characterized immediately after deposition or annealing.

*Characterization of Structural and Physical Properties:* The sample structure was characterized with a high-resolution four-circle XRD. Temperature-dependent *dc* transport measurements were conducted using the van der Pauw geometry with a 14 T Physical Property Measurement System (PPMS). Optical spectroscopy was performed using a spectroscopic ellipsometer between 1.25 and 5.00 eV at an incident angle of 70º. A simple two-layer model (film/substrate) was used to extract dielectric functions and optical conductivity. Valence state and oxygen stoichiometry via

XAS were performed at the beamline 4-ID-C of the Advanced Photon Source at Argonne National Laboratory.

*DFT Calculations:* All modeling calculations have been performed within density functional theory (DFT) employing the Vienna Ab-initio Simulations Package (VASP) code. We have used $2 \times 2 \times 1$ supercells for all calculations containing 144 atoms. Projector-augmented wave pseudopotentials have been used with an energy cut of 600 eV. The activation energy ($E_a$) barriers ($\Delta E_a$) for oxygen ions and the intermediate transition states have been computed using the Nudged Elastic Band (NEB) method as implemented in the VASP code. The energy barriers have been optimized until the forces on each image were converged to 0.004 eV/Å. In order to account for strong correlations, the cobalt *d* orbitals are treated within the local spin density (LSD) approximation with Hubbard *U* corrections. A *U* value of 7.5 eV was chosen, the electronic structure of which matched closely to those computed with the Hybrid Scuzeria Ernzerhof (HSE) functional. Further details can be found in Supplementary Information and in previous work.[26]

**Supporting Information**
Supporting Information is available from the Wiley Online Library or from the author.

**Acknowledgements**
This work was supported by the U.S. Department of Energy, Office of Science, Basic Energy Sciences, Materials Science and Engineering Division. Use of the Advanced Photon Source was supported by the U. S. Department of Energy, Office of Science, under Contract No.DE-AC02-06CH11357.

Received: ((will be filled in by the editorial staff))
Revised: ((will be filled in by the editorial staff))
Published online: ((will be filled in by the editorial staff

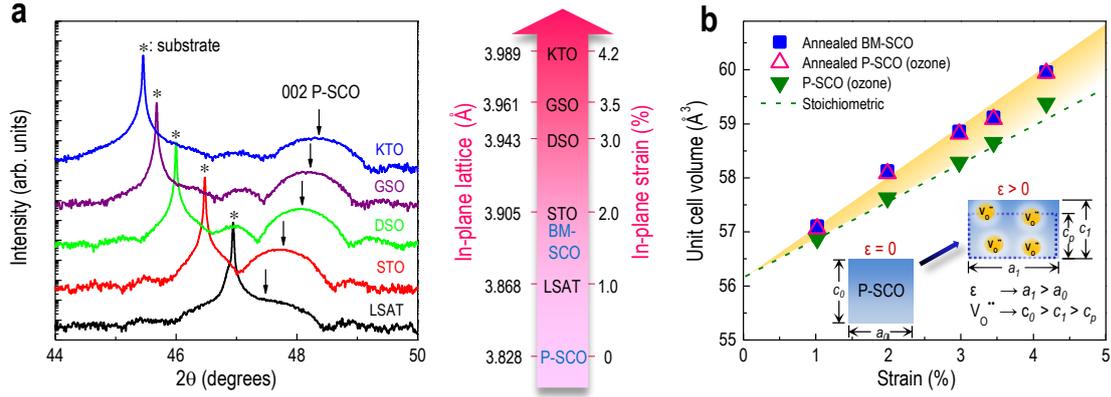

**Figure 1. Strain control of oxygen stoichiometry in epitaxial P-SCO. a)** XRD *θ-2θ* scans around the 002 peak of topotactically oxidized P-SCO films on different substrates. Substrate peaks are noted by asterisks (*). In-plane lattice parameters and biaxial strain applied to P-SCO are provided adjacent to the scan. **b)** As the in-plane parameter $a_0$ is stretched to $a_1$ due to tensile strain, there is the expected shift in the 002 peak from $c_0$ to the contracted $c_1$ out-of-plane parameter. However, $c_1$ is larger than would be expected from a Poisson contraction ($c_p$ and dotted box) of stoichiometric P-SCO. To deconvolute the effects of strain and vacancy formation, unit cell volumes as a function of biaxial strain for either annealed BM-SCO or P-SCO$_{ozone}$ films are given. The similarity between the annealed films indicates an equilibrium oxygen stoichiometry attained through either the deintercalation or intercalation of oxygen. The lattice volume sizes are compared to the ones of fully oxidized SrCoO$_x$ ($x \geq 2.9$) (P-SCO$_{ozone}$) films from ozone growth, which revealed a Poisson ratio $v = 0.26$ (dashed line). As denoted by the shaded region, the higher the tensile strain, the higher the volume deviation from the dashed line, indicating an increase in oxygen vacancies.

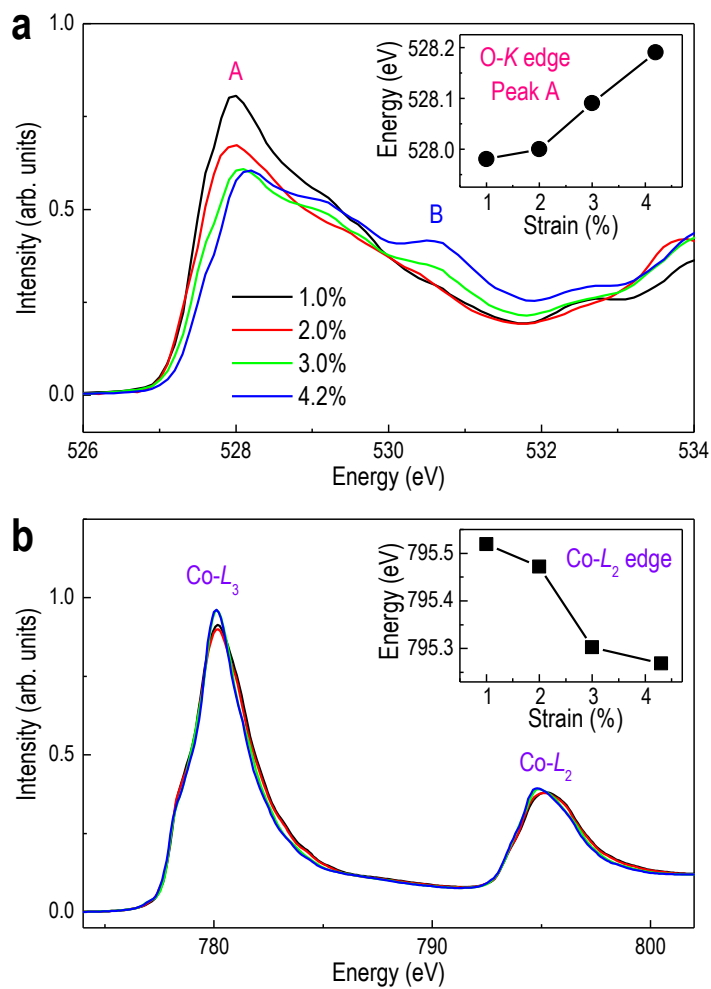

**Figure 2. Evidence for preferential oxygen loss in tensile strained P-SCO. a)** XAS O-$K$ edge of annealed P-SCO films on LSAT ($\varepsilon = 1.0\%$) through KTO ($\varepsilon = 4.2\%$) substrates. Peaks at ~528 eV (A) and ~530.5 eV (B) correspond to Co3$d$-O2$p$ hybridization, respectively associated with either fully-oxidized or partially-oxidized Co coordination. **b)** The Co-$L_{2,3}$ edges indicate the shift in Co valency from $Co^{4+}$ to $Co^{3+}$ as vacancies are induced by tensile strain.

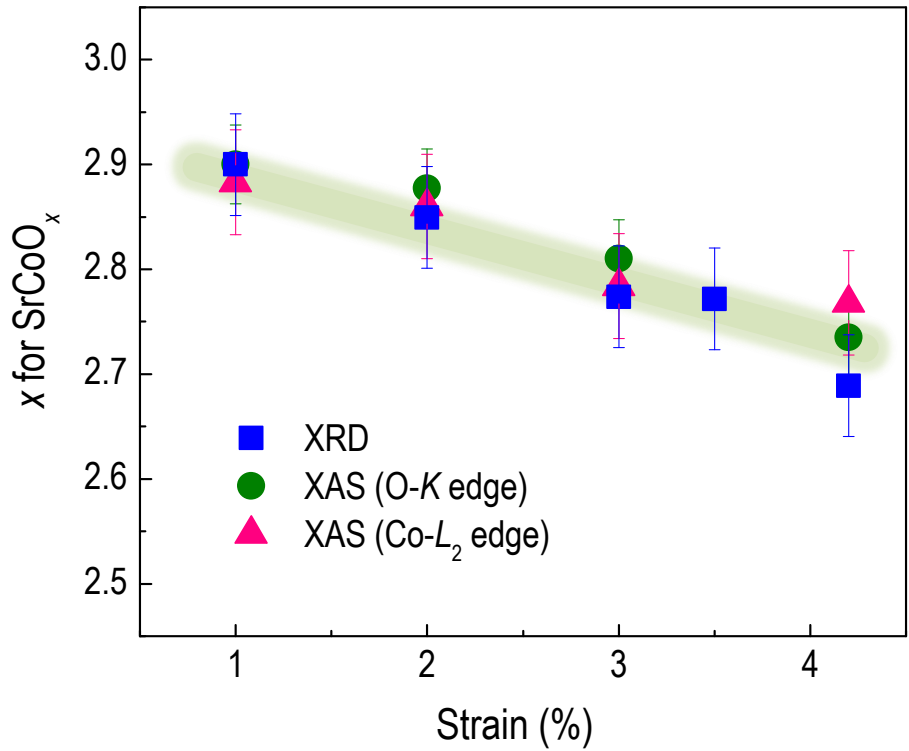

**Figure 3. Oxygen deficiency in tensile strained P-SCO. a)** Oxygen non-stoichiometry $x$ in $SrCoO_x$ determined by comparing XRD-based volume changes and peak position shifts of XAS O-$K$ and Co-$L_2$ edges to the literature values.[29, 30, 34, 37] From this combined data set, $x$ is found to range from just below 3 to 2.75 as tensile strain increases from $\varepsilon = 1\%$ to $\varepsilon = 4.2\%$

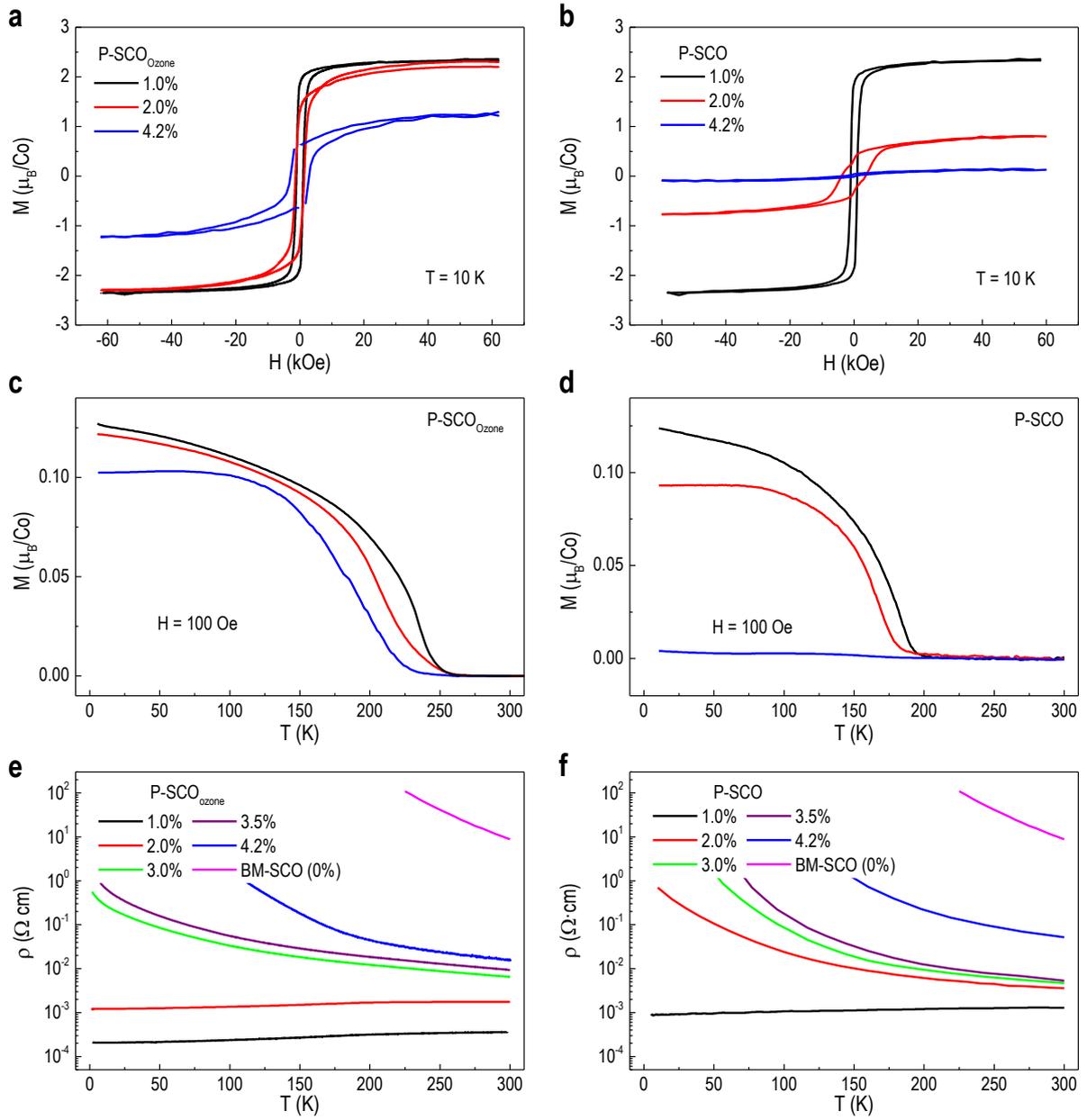

**Figure 4. Strain dependent physical properties.** The magnetization and electrical transport of P-SCO$_{ozone}$ are shown on the left panels in a), c), and e) while those of annealed P-SCO are displayed on the right panels in b), d), and e). We note that the growth in highly oxidizing ambient using ozone could improve the oxygen stoichiometry. Tensile strain results in deterioration of the ferromagnetic metallic ground state, and the addition of strain-induced changes in oxygen stoichiometry radically drives the P-SCO towards becoming an antiferromagnetic insulator. As grown BM-SCO on STO is included as a reference, putting an upper boundary on the oxygen deficiency for P-SCO.

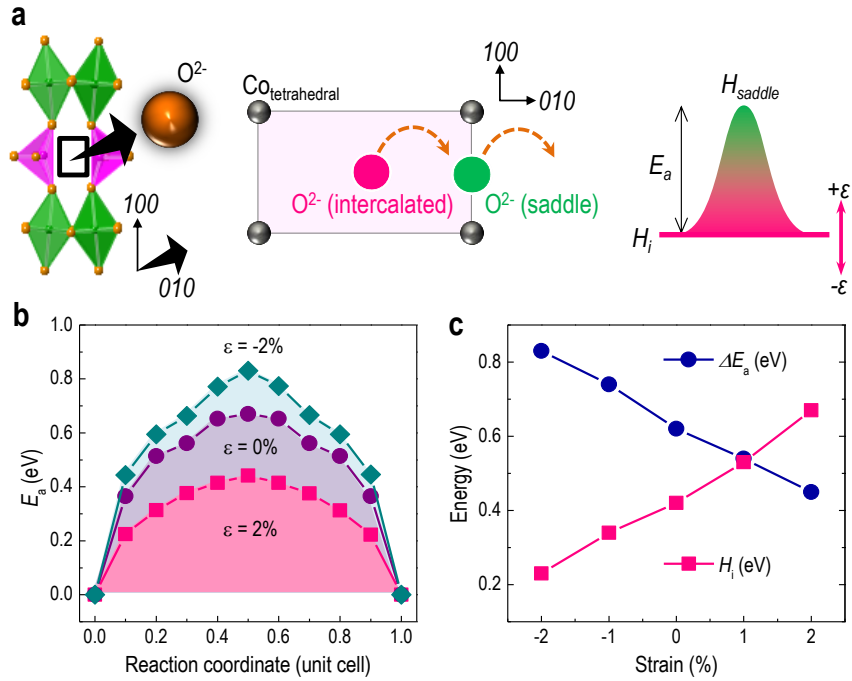

**Figure 5. Strain dependent oxygen activation energies. a)** A schematic indicating the motion and energy barriers for oxygen ion movement in BM-SCO through the OVCs when undergoing a topotactic oxidation to P-SCO. The OVC channel is within the tetrahedral (pink) layer, which is bounded by the octahedral layers (green). While the intercalation enthalpy, $H_i$, at the vacancy sites changes significantly with strain, the point of highest energy, the saddle point ($H_{saddle}$), is not considerably affected. **b)** The difference between the two, the activation energy ($E_a$), as an oxygen ion is moved from one intercalation site to another along the [010] direction of the OVC for -2, 0, and 2% strain states. **c)** A summary of the activation energy barrier ($\Delta E_a$) and intercalation enthalpy ($H_i$) as a function of strain, showing that tensile strain facilitates oxygen movement through the OVCs and reduces the stability of oxygen intercalation (leading to vacancy formation), whereas compressive strain preferentially immobilizes oxygen.

# Supporting Information

**Strain Control of Oxygen Vacancies in Epitaxial Strontium Cobaltite Films**

*Jonathan R. Petrie[1], Chandrima Mitra[1], Hyoungjeen Jeen[1], Woo Seok Choi[1], Tricia L. Meyer[1], Fernando A. Reboredo[1], John W. Freeland[2], Gyula Eres[1], and Ho Nyung Lee[1*]*

**Overview of DFT and Enthalpy Calculations**

Calculations can be based off previous work.[26] To better understand the site stability in BM-SCO, we first note that the enthalpy minima correspond to the intercalating bonding sites (Co-O bond) within the OVCs of the open framework structure. In diffusing from one bonding site to the other along the [010] direction in the channel, $O^{2-}$ crosses an intermediate saddle point. In principle, the difference in formation enthalpy between the intercalation site and saddle point (or $H_i$ and $H_{saddle}$) could also yield the activation energy barrier, $\Delta E_a$. Since NEB calculations can only provide an estimate of the difference ($\Delta E_a$) but not the actual formation enthalpies at these two points, we employ first-principles calculations to compute $H$ as a function of strain. We adopt the Zhang-Northrup formalism, which is the standard approach for computing defect formation enthalpies in first-principles calculations.[48] Within this approach, the formation enthalpy of a defect X is formulated as:

$$H^f[X] = E_{tot}[X] - E_{tot}[bulk] - \sum_k n_k \mu_k \qquad (1)$$

The first and second terms on the right-hand side of *Eq.* (1) refer to the total energies of the supercell with and without the defect, respectively, $n_k$ the number of atoms of type k that have been added ($n_k > 0$) or removed ($n_k < 0$) from the supercell in order to create the defect, and $\mu_k$ are the corresponding chemical potentials. We find that, for tensile strain, $H_i$ increases, suggesting less stability of $O^{2-}$ on the intercalating bonding site due to a lengthening of the Co-O bond and subsequent decrease in stabilizing Co 3*d*-O 2*p* hybridization. The opposite effect occurs for compressive strain as $H_i$ decreases. In principle, one could do a similar thing for the saddle point; however, tensile strain results in only a slight lowering of $H_{saddle}$. The opposite happens for compressive strain. The net effect is, therefore, a reduction and enlargement in the overall value of $\Delta E_a$ with tensile and compressive strain, respectively.

As experimental observations suggest, oxygen stoichiometry in BM-SCO changes with the application of strain. This is verified by our theoretical calculations of $O^{2-}$ intercalation formation enthalpy, $H_i$, as a function of strain (see Fig. 3b). The probability of occurrence, $P$, of an intercalated oxygen, and hence the variation of oxygen stoichiometry is directly related to $H_i$ as:

$$P = \frac{e^{-H_i/KT}}{1+ e^{-H_i/KT}} \qquad (2)$$

The concentration, $c$, of intercalated oxygen (in equilibrium), is related to $P$ and hence $H_i$ as:

$$c = N_{sites}P \qquad (3)$$

$N_{sites}$ refer the number of available sites in the lattice (per unit volume). According to our calculations $H_i$ in an unstrained BM-SCO has a value of 0.42 eV when the reference chemical potential of oxygen is taken as half of the total energy of an oxygen molecule (this assumes oxygen rich condition when the system is in equilibrium with the oxygen reservoir at 1 atm pressure and 298 K). As can be seen from Fig. 3b, the application of a 2% compressive strain reduces $H_i$ to 0.2 eV which would mean an increase in oxygen concentration by almost an order of magnitude according to equation (3). Note that the equilibrium lattice constant of perovskite $SrCoO_3$ is less than BM-SCO and hence compressing the BM-SCO lattice is expected to stabilize filling up the vacancy channels in its open framework with oxygen. This is consistent with our calculations as well. Tensile strain on the other hand increases $H_i$, which means that the vacancy channels would no longer be easily filled up.

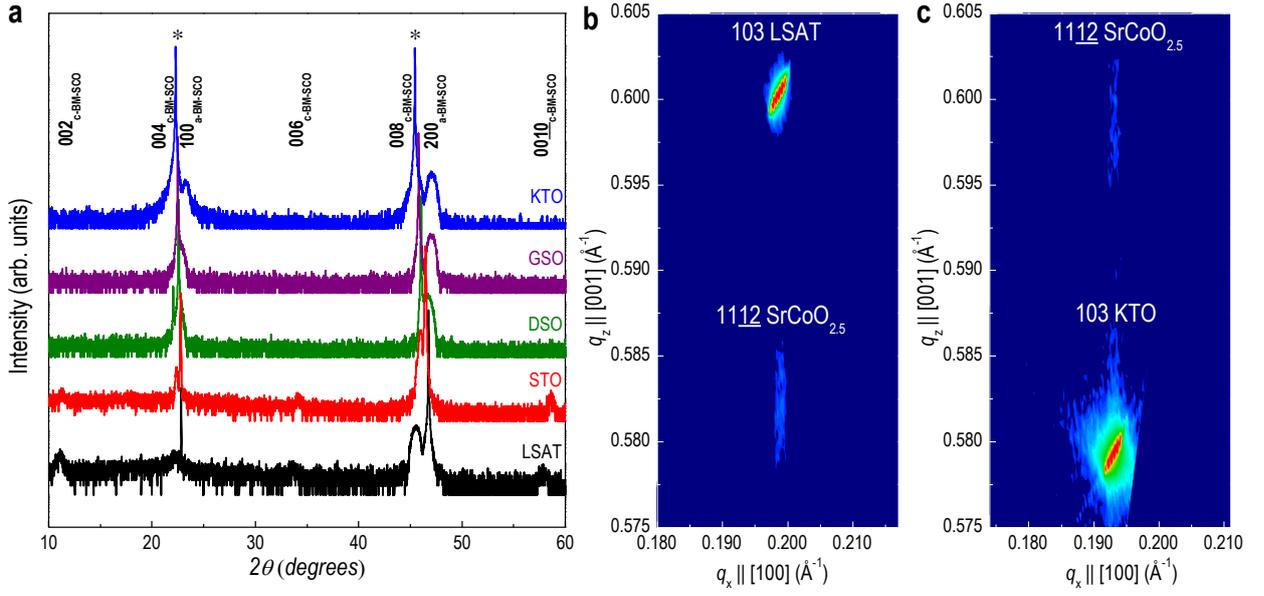

**Figure S1. Coherent growth of SrCoO$_{2.5}$ on various substrates. a)** XRD $\theta$-$2\theta$ scans of as-deposited BM-SCO films on various substrates. Due to the better lattice mismatch along the *c*-axis of BM-SCO, the films on DSO, GSO, and KTO are *a*-axis-oriented, while the rest films on LSAT and STO are *c*-axis oriented. Substrate peaks are noted by asterisks (*). **b)** RSMs of LSAT ($\varepsilon = 1.0\%$) and KTO ($\varepsilon = 4.2\%$) are shown to confirm the coherent growth on the least and most tensile-strained substrates, respectively.

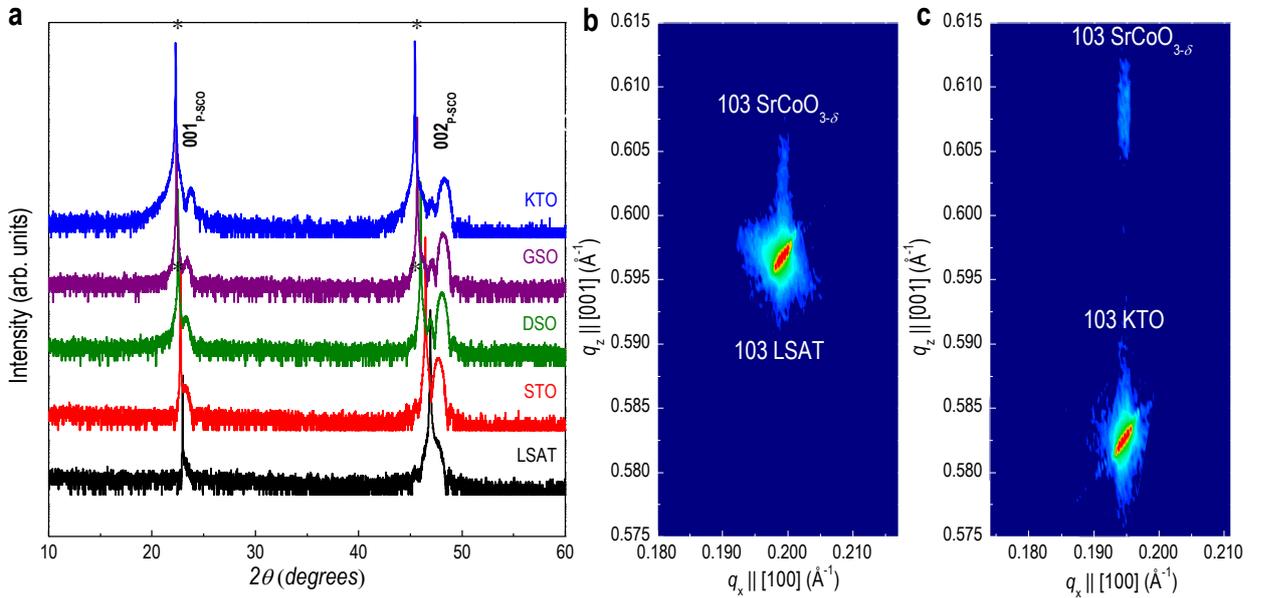

**Figure S2. Topotactically oxidized SrCoO$_{3-\delta}$ films. a)** XRD $\theta$-$2\theta$ scans of topotactically oxidized P-SCO films on lattice-mismatched substrates. The lack of half-order peaks and the shift in the 002 peak are indicative of the perovskite SrCoO$_{3-\delta}$ (P-SCO), where $\delta \leq 0.25$. Substrate peaks are marked by asterisks (*). **b)** RSMs of P-SCO films on LSAT and KTO are shown to confirm coherent growth on the least and most tensile-strained substrates, respectively.

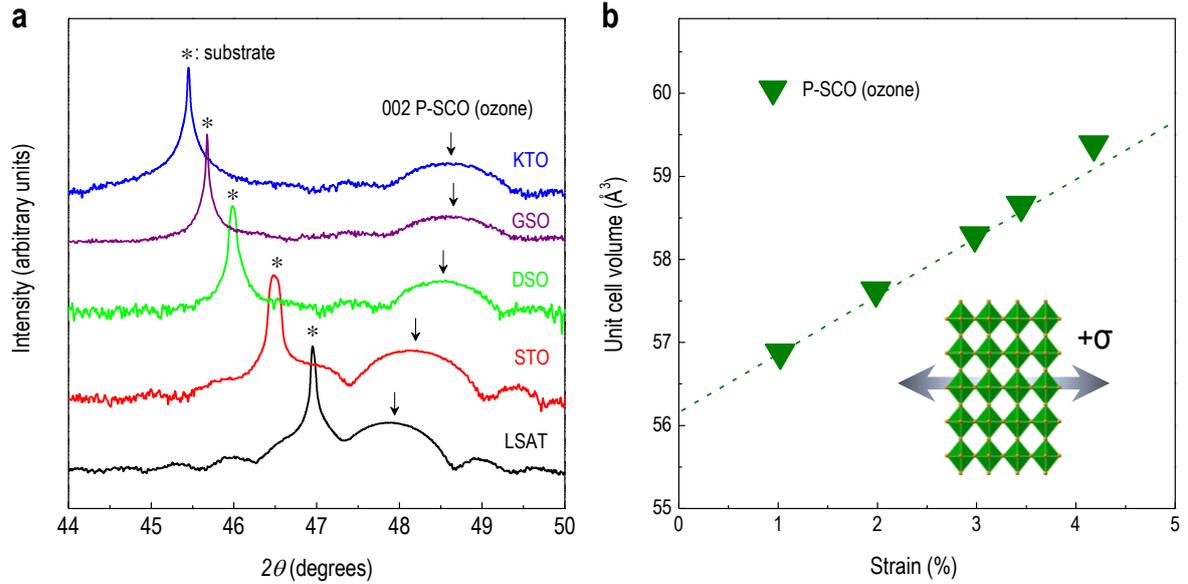

**Figure S3. Growth of P-SCO$_{ozone}$ in highly activated oxygen. a)** XRD $\theta$-$2\theta$ scans around the 002 peak of ozone-grown P-SCO (P-SCO$_{ozone}$) films on different substrates. Substrate peaks are noted by asterisks (*) **b)** As the in-plane parameter is stretched, there is the expected shift in the 002 peak from contraction of the out-of-plane parameter, fitting a Poisson ratio $\nu = 0.26$ (dashed line).

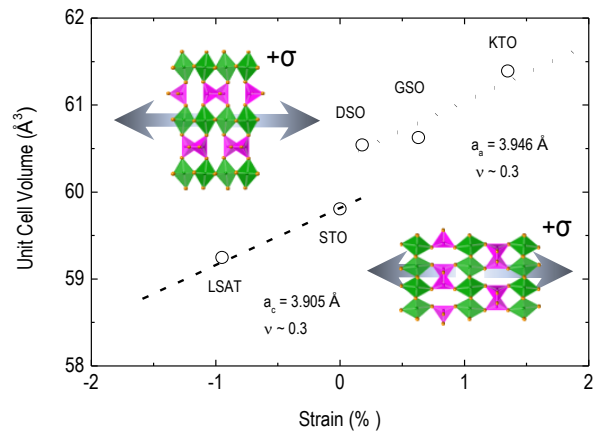

**Figure S4. Pseudocubic unit cell volume as a function of biaxial strain for as-grown BM-SCO.** Bulk values derived from a Poisson ratio $\nu = 0.30$ are indicated by a dashed line for the c-axis and dotted line for the a-axis oriented BM-SCO. The BM-SCO films show no significant deviation from strained bulk BM-SCO.

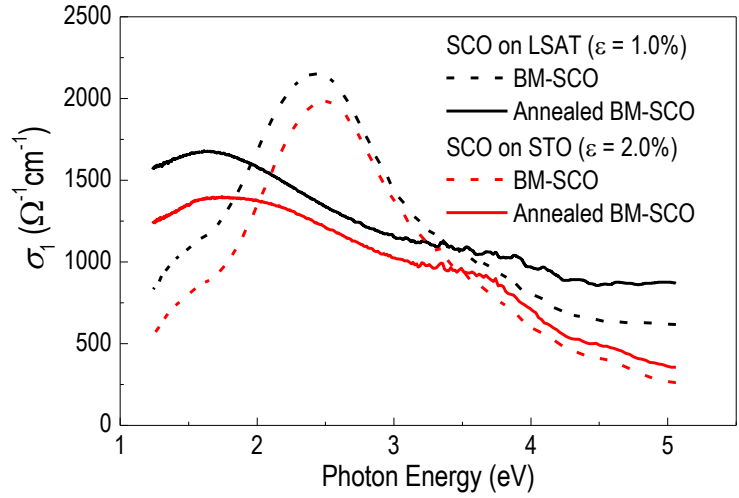

**Figure S5. Comparison of the optical properties between BM-SCO and P-SCO.** Representative optical conductivity spectra of BM-SCO films and topotactically oxidized P-SCO films on LSAT ($\varepsilon = 1.0\%$) and STO ($\varepsilon = 2.0\%$) substrates. A clear difference between the two phases is observed, i.e. a clear Drude feature for P-SCO and a clear insulating ground state with a low band gap for BM-SCO.